\documentstyle[prl,aps,epsf,twocolumn]{revtex}
\begin{document}
\draft
\tighten
\title{
Localization of Elastic Layers by Correlated Disorder }
\author{Leon Balents}
\address{Department of Physics, Harvard University, Cambridge, MA 02138}

\date{\today}
\maketitle

\widetext

\begin{abstract}
The equilibrium behavior of a system of elastic layers under tension
in the presence of correlated disorder is studied using functional
renormalization group techniques.  The model exhibits many of the
features of the Bose glass phase of type II superconductors induced by
columnar defects, but may be more directly applicable to charge density
waves, incommensurate striped magnetic phases, stacked membranes under
tension, vicinal crystal surfaces, or superconducting
``vortex--chains''.  Below five dimensions, an epsilon expansion for
the stable zero temperature fixed point yields the properties of the
glassy phase.  Transverse to the direction of correlation, the
randomness induces logarithmic growth of displacements.  Displacements
are strongly localized in the correlation direction.  The absence of a
response to a weak applied transverse field (transverse Meissner
effect) is demonstrated analytically.  In this simple model, the
localized phase is stable to point disorder, in contrast to the
behavior in the presence of dislocations, in which the converse is
believed to be true.
\end{abstract}

\narrowtext

Recently, considerable progress has been made in understanding the
behavior of elastic media in the presence of randomness.  Examples include
single flux lines in a dirty superconductor \cite{KZ,NL}, interfaces in
random magnets \cite{DSFFRG,BFmanifold,HH}, charge density waves
\cite{NFcdw}, and the vortex glass phase of bulk
superconductors \cite{VG,FGLV}.  The experimental work of Civale et. al.
\cite{Civale} has demonstrated the feasibility of creating
superconducting samples with {\sl correlated} (columnar) 
\begin{figure}
\epsfxsize=3.0truein
\hskip 0.5truein \epsffile{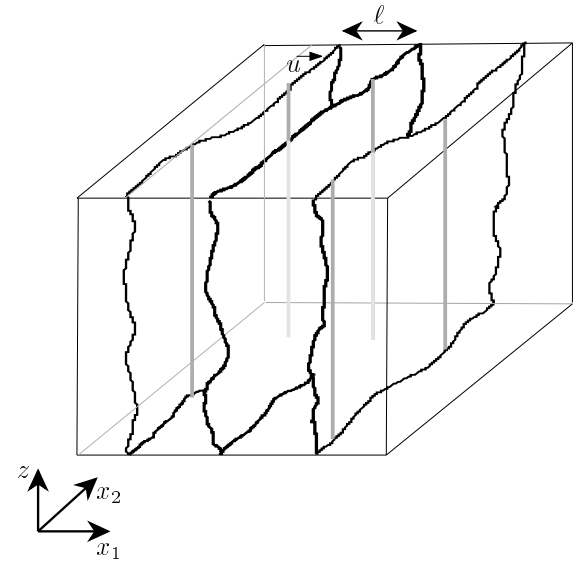}
\caption{A stack of layers in three dimensions fluctuating in the
presence of correlated disorder.  Dark and light thick vertical lines
indicate individual correlated pins, in the foreground and hidden from
view, respectively.  The spacing along the layering ($x_1$) axis,
$\ell$, and the displacement field $u({\bf x},z)$, are indicated.}
\label{layerfig}
\end{figure}
\noindent disorder.  One of the most striking aspects of the resulting
localized Bose glass phase\cite{NV,Lyuk}\ is the existence of  
\newline\vskip 1.9truein\noindent
a finite critical mismatch angle $\theta_c$ between the applied field and the
correlation direction, such that the flux lines remain parallel to the
correlation axis for $\theta <\theta_c$.  In this paper, an analogous
localized phase is studied analytically near 5 dimensions for a
somewhat simplified model which may describe other systems of
interest.  The pulling--away from the correlation axis for $\theta >
\theta_c$ is discussed in Ref.\cite{BRinprep}.

Consider a system made up of a stack of layers (for general $d$, these
will be oriented manifolds of dimension $d-1$, for instance
interfaces) which may fluctuate in the perpendicular direction, but
may not pass through one another (see Fig.\ref{layerfig}).  Such a
model could describe a charge density wave, the domain walls of an
incommensurate striped phase in a magnet \cite{VillainBak}, a stack of
membranes under tension, or, for $d=2$, a set of steps on a miscut
crystal surface\cite{BKrough}.  Another possibile realization is
suggested by recent observations of ``vortex--chains'' in
YBCO\cite{vortexchain}, which should fluctuate as layers defined by
the chain and magnetic field directions.  The displacement of the
$k^{th}$ layer, $u_k$, is defined by $x_{k,1} \equiv k\ell +
u_k(x_2,\ldots,x_{d-1},z)$, where $\ell$ is the average layer
separation, and $x_1$ is taken to be the layering axis.  The last
coordinate $x_d \equiv z$ has been distinguished as the direction of
correlations for the random potential.  The use of a displacement
field neglects dislocations, which will be important in many systems
beyond some length scale.  Taking the continuum limit
$u_k(x_2,\ldots,x_{d-1},z)
\rightarrow u({\bf x},z)$, the Hamiltonian is
\begin{eqnarray}
H & =  \int\! d^{d-1}\!{\bf x}dz \Big\{ & {K \over 2}\left( \nabla u
\right)^2 + {\tilde{K} \over 2} \left( \partial_z u \right)^2 +
h\partial_z u \nonumber \\ 
& &  + V_C(u,{\bf x}) + V_P(u,{\bf x},z) \Big\},
\label{hamiltonian}
\end{eqnarray}
where the $x_1$ coordinate has been rescaled to remove the anisotropy
of the in-layer elasticity, and a momentum cut-off $\Lambda \sim
1/\ell$ is implicitly included in the ${\bf x}$ direction (but not in
$z$) due to the discreteness of the layers.  A non-zero $h$ represents
a force tending to tilt the layers, such as that caused by a change in
the applied field in a superconductor, or by tilted boundary
conditions.  Only the response to a small $h$ will be considered
here.  $V_C$ and $V_P$ are random potentials describing, respectively,
columnar and point disorder.  By including both types of pinning, it
is possible to study the competition between vortex-glass and
Bose-glass like phases in this dislocation-free model\cite{BKFL}.
Because of the periodicity of the stack of layers, these potentials
must be periodic functions of $u$.  For weak disorder, the fixed point
potentials will have Gaussian distributions, with the correlation
functions
\begin{eqnarray}
\langle V_C(\!u,\!{\bf x})V_C(\!u',\!{\bf x}') \rangle
 & = & R_C(\!u\!-\!u')\delta({\bf
\!x\!-\!x'}), \label{cfdefcorr} \\
\langle V_P(u,\!{\bf x},\!z)V_P(\!u',\!{\bf x}',\!z')
\rangle 
& = & R_P(\!u\!-\!u')\delta({\bf
\! x\!-\!x'})\delta(\!z\!-\!z').
\label{cfdefpt}
\end{eqnarray}
Choosing $\ell = 2\pi$, periodicity implies $R_{C,P}(u+2\pi) =
R_{C,P}(u)$.

To understand the behavior on long wavelengths, a renormalization
group (RG) analysis was performed using the methods of
Ref.\cite{BFmanifold}.  Lengths are rescaled parallel and
perpendicular to the disorder, according to ${\bf x} \rightarrow b{\bf
x}$ and $z \rightarrow b^\chi z$.  The displacement cannot be rescaled
due to periodicity.  The cut-off is kept fixed by integrating out
modes with ${\bf x}$ momenta in a shell $\Lambda/b < p < \Lambda$
(momenta in the $z$ direction are unrestricted).  Formally, this
procedure can be carried out by expanding the partition function in
$V_C$ and $V_P$ and performing the functional integrals order by
order.  The resulting terms are functions of the remaining modes, and
upon reexponentiation yield flows of the interactions in the
Hamiltonian.  Since $u$ is dimensionless, however, $R_C$ and $R_P$
cannot be taken to have the $u^4$ form commonly encountered in RG
studies of phase transitions.

For $d=5-\epsilon$, the behavior of the system is dominated by a
non-trivial zero-temperature fixed point.  The scale changes yield the
eigenvalues $\lambda_T \equiv -\theta = -2 - \chi +O(\epsilon)$
and $\lambda_{R_P} = - \chi + O(\epsilon)$.  Naively, $\chi = 1 +
O(\epsilon)$, but this point will be returned to later.  Regardless of
the value of $\chi$ (so long as it is positive), $T$ and $R_P$ are
formally irrelevant, so I will begin by working directly at
$T=R_P=0$.

At zero temperature, the computation of the partition function reduces
to the optimization of the hamiltonian.  At each step of the RG, the
lowest energy configuration of the modes in the shell is found as a
function of the low momentum modes, which are held fixed (see,
e.g. Ref.\cite{BFmanifold}).  The minimum of H (Eq.\ref{hamiltonian})
clearly satisfies $\partial_z u = 0$ {\sl exactly}.  The
renormalization of $R_C$, therefore, must be identical to the case of
point disorder in $d-1$ dimensions.  In addition, the statistical
Galilean invariance of the $d-1$ dimensional model \cite{galilean}
guarantees that temperature is only trivially renormalized by the
scale changes.  There is, however, a non-trivial renormalization of
$\tilde{K}$, since the iterative minimization does not throw out
excitation information until it is on a scale smaller than the
cut-off.  Defining the force-force correlation function $\Delta_{C}(u)
\equiv -R_{C}^{''}(u)$, the RG equations for $b=e^{dl}$ to lowest
non-trivial order are
\begin{eqnarray}
{{d \tilde{K}} \over {d l}} & = & \left( 2 -
2\chi - \Delta^{''}_{C}(0)/(8\pi^2)\right)\tilde{K},
\label{RGsimple} \\
{{d \Delta_C} \over {dl}} & = & \epsilon\Delta_C - {1 \over
{8\pi^2}} \left[ \left(\Delta'_C\right)^2 +
\Delta_C^{''}\left(\Delta_C-\Delta_C(0)\right)\right], \label{RGdelta}
\end{eqnarray}
where, for simplicity, I have taken $K = 1$.  Eq.\ref{RGsimple}\ is
obtained assuming analyticity of $\Delta_C$, and will be corrected
later.  Eq.\ref{RGdelta} has been derived previously by a number of
other authors \cite{DSFanisotropies,NFcdw,LGflux}, and has the stable
$2\pi$-periodic fixed point solution
\begin{equation}
\Delta_C(u) = {{4\pi^2\epsilon} \over 3}\left[(u-\pi)^2 -
\pi^2/3\right], \qquad {\rm for } \; 0<u<2\pi \; .
\label{FPfunction}
\end{equation}
This fixed point solution leads\cite{LGflux}\ to logarithmic displacement
fluctuations.   First order perturbation theory in $R_C$, evaluated at
the fixed point (Eq.\ref{FPfunction}), gives the $O(\epsilon)$ result
\begin{equation}
\left\langle \left( \bar{u}({\bf x}) - \bar{u}({\bf 0}) \right)^2
\right\rangle \sim {{2\pi^2\epsilon} \over 9}\ln |{\bf x}|,
\end{equation}
where $\bar{u}({\bf x})$ is defined as the $z$ average of $u({\bf
x},z)$, and the angular brackets denote both thermal and disorder
averaging.  The fixed point function has a slope discontinuity when
$u=0$, so that $\Delta^{''}_C(0) = -\infty$! Note that the procedure
employed in the near-threshold dynamic problem of Ref.\cite{NFcdw} for
handling this singularity does not apply here due to the different
physics of force--free equilibrium.  The divergence implies that the
feedback of $\Delta_C(u)$ to the elastic terms (Eq.\ref{RGsimple})
must be re-analyzed.  With no assumptions on the analyticity of
$\Delta_C(u)$, the term generated by the RG which led to this equation
takes the form
\begin{eqnarray}
\left.{{dH} \over {dl}}\right|_{\rm elastic} & = & - {1
\over 2} {\Lambda^3 \over {16\pi^2\tilde{K}^{1/2}}}\int\!\!
d^{d\!-\!1}\!{\bf x}d\!zd\!z' \Delta_C(\!u(z) \!-\! u(z')) \nonumber \\ 
& & \times \exp\left[ - {\Lambda \over
\tilde{K}^{1/2}} |z-z'|\right].
\label{generated}
\end{eqnarray}
The exponential decay for large separations justifies a gradient
expansion of $u(z)$ near $z'$, which combined with the small $u$
behavior $\Delta_C(u) \approx {{8\pi^2\epsilon} \over 3}(\pi^2/3 -
\pi|u| + u^2/2)$ yields a new term in the Hamiltonian
\begin{equation}
\Delta H = {\sigma \over 2} \int\!\! d^{d\!-\!1}\!{\bf x}\,dz
|\partial_z u|.
\label{absvaluestiffness}
\end{equation}
Eq.\ref{RGsimple}\ is replaced by the pair of equations
\begin{eqnarray}
{{d\sigma} \over {dl}} & = & (2-\chi)\sigma +
{{\pi\Lambda\epsilon\tilde{K}^{1/2}}
\over 3}, \label{sigmarenorm} \\
{{d\tilde{K}} \over {dl}} & = & \left( 2 -
2\chi - \epsilon/3)\right)\tilde{K} \label{RGcorrect},
\end{eqnarray}

The new elastic term has a direct physical meaning.  Consider the
response of the system to an infinitesimal field $h$.  In the absence
of disorder, the minimum of the Hamiltonian as a function of
$\partial_z u$ is shifted over by an amount linear in $h$, resulting
in a response $\theta \equiv \partial_z u \propto h$.  This will be
true regardless of the magnitude of $\tilde{K}$.  If the elastic term
has the form of $\sigma|\partial_z u|$, however, the minimum will
remain at $\partial_z u = 0$ for all $h<K$.  This implies the
existence of a finite threshold force, below which the layers remain
locked in their localized positions.

It is clear, both from the physical interpretation above and from the
form of Eq.\ref{sigmarenorm} and Eq.\ref{RGcorrect}, that the simple
expectation of $O(\epsilon)$ corrections to $\chi$ will not hold.
Eq.\ref{sigmarenorm} suggests a value of $\chi=2$, in accord with a
simple ``random walk'' picture of excursions from the localized ground
state.  A careful analysis, however, must continue the RG procedure
after the new elastic term has been generated.  While this cannot
affect the renormalization of $\Delta_C$, it will probably affect the
$z$-dependent portions of the Hamiltonian.  It is not known whether a
consistent treatment of such a non-analytic elastic term is possible
even to $O(\epsilon)$.  Justification of the value $\chi=2$, even to
lowest order, requires a detailed investigation of the zero
temperature RG along the lines of that performed in appendix C of
Ref.\cite{BFmanifold}.  Such an analysis is not available at present.

Because the vanishing of the tilt response requires $\Delta^{''}_C(0)
= - \infty$, the weak field behavior is sensitive to corrections from
formally irrelevant operators such as temperature and point disorder.
To safely conclude the stability of the correlated phase, the
corrections will now be studied in more detail. These operators yield
additional renormalizations of $\Delta_C$
\begin{equation}
\left.{{d\Delta_C} \over {dl}} \right|_{T,\Delta_P} = {\Lambda \over
{16\pi^2\tilde{K}^{1/2}}}\left[ \Lambda^2 T +
{1 \over 2}\Delta_P(0) \right]\Delta^{''}_C.
\label{irrelfeedback}
\end{equation}
Because of the presence of such terms, analytic behavior of
$\Delta_C(u)$ will persist within a narrow boundary layer around
$u=0$.  Since temperature and uncorrelated disorder are irrelevant
operators, the width $w$ of the boundary layer decreases under the RG
(note that since $\theta > |\lambda_{R_P}|$, point disorder will
dominate this width at long length scales).  Although this rounding is
not a property of the fixed point value of $\Delta_C$, the rapid
divergence of $\Delta^{''}_C(u)$ in the absence of the irrelevant
operators (see, e.g. Ref.\cite{DSFFRG}) indicates that the size of the
smoothed region is determined by the terms in Eq.\ref{irrelfeedback}.

As the RG iterates to longer length scales, $\Delta_C(u)$ sharpens up
near the origin.  From Eq.\ref{FPfunction}, the jump in slope across
the boundary layer, where $\Delta_C(u)$ must match its fixed point
value, is $O(\epsilon)$.  Since this change in slope must be
accommodated over a width $w$, the curvature $\Delta^{''}_C(0) \sim
-\epsilon/w$.  Equating the terms in Eq.\ref{irrelfeedback} to those in
Eq.\ref{RGdelta} gives the scaling of the boundary layer width
\begin{equation}
w(l) \sim {\Lambda \over {\epsilon\tilde{K}^{1/2}}}\left[
\Lambda^2 T(l) + {1 \over 2} \Delta_P(0;l) \right].
\label{wscale}
\end{equation}

\begin{figure}
\epsfxsize=3.0truein
\hskip 0.5truein \epsffile{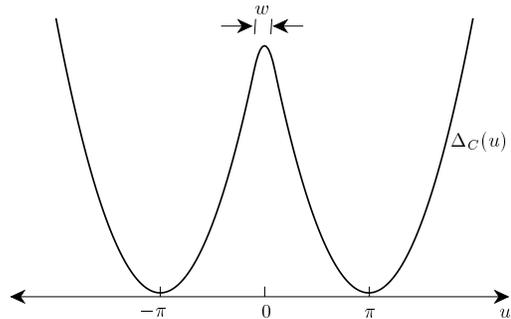}
\caption{A schematic illustration of the columnar disorder correlation
function $\Delta_C(u)$.  For $|u-2\pi k| \gg w$, $\Delta_C(u)$ has the
simple form of Eq.6.  For $|u-2\pi k| \ll w$, $\Delta_C(u)$ is
rounded, with curvature of $O(\epsilon/w)$. }
\label{corrfnfig}
\end{figure}

$\Delta_C(u)$ is illustrated in Fig.\ref{corrfnfig}.  From the structure of
Eq.\ref{generated}, it is clear that (for small gradients) the
generated terms take the form
\begin{equation}
\left.{{d H} \over {dl}}\right|_{\rm elastic} = \epsilon\Lambda^2 \int\!
d^{d\!-\!1}{\bf x} dz \; f\left[{\tilde{K}^{1/2} \over \Lambda}
|\partial_z u| \right],
\label{functionofgradient}
\end{equation}
where $f(u) \sim |u|$ for $w \ll |u| \ll 1$, but is smooth for $|u|
\ll w$.  

Tilting for small $h$ can only occur for $\theta$ within the smooth
boundary layer.  Although Eq.\ref{functionofgradient}\ is valid only
while feedback from $\sigma$ can be neglected, it does demonstrate
that this boundary layer width, $w$, vanishes exponentially as a
function of length scale.  Since the average tilt $\theta$ is a bulk
($q=0$) quantity, the boundary layer does not contribute.
By contrast, it is precisely the small boundary-layer in
Eq.\ref{functionofgradient}\ which leads to thermal creep in weakly
driven interfaces and charge-density waves \cite{DSFunpub}.  The only
important effect of the interaction between point and columnar
disorder in this model is to {\sl decrease} the relevance of point
disorder, both through an increase in $\chi$ and through terms from
$\Delta_C$ feeding into the RG equation for $\Delta_P$ (not shown here).

In conclusion, the system of elastic layers studied here forms a
disordered-dominated phase in the presence of correlated randomness.
Despite the reduction in the number of components of the displacement
field and the neglect of dislocations, the model exhibits {\sl
analytically} many of the properties of the Bose glass.  The physical
properties are governed by a zero temperature fixed point, at which
the layers are completely parallel to the direction of correlation.
Due to this localization, the tilt response to an applied transverse
field vanishes below some finite threshold field, corresponding to the
``transverse Meissner effect'' of Ref.\cite{NV}.  The fluctuations of
the layers perpendicular to the correlations grow logarithmically,
with a universal prefactor.  In contrast to the Bose glass case
\cite{HN}, the system is stable against uncorrelated disorder.
It thus appears that dislocations are necessary to make point disorder
relevant in the localized phase.  Naive arguments suggest a simple
``random walk'' scaling, $\chi=2$, for the low-lying excitations,
although it is quite likely that this result will be corrected by
non-trivial renormalizations.  The analysis of these excursions from the
ground state and the extension to more components remain interesting
open problems.  

It is a pleasure to acknowledge discussions with Daniel Fisher, Leo
Radzihovsky, Randall Kamien, Mehran Kardar, and Thomas Natterman.
This research was supported by the NSF through grant number
DMR--90--01519, and Harvard University's Materials Research
Lab, grant DMR--91--06237.

\end{document}